\documentstyle[A4,12pt]{article}

\begin{document}

\title{On Stress Analysis for Cracks in Elastic Materials with Voids}
\author{M.\,Ciarletta, \quad G.\,Iovane \\
D.I.I.M.A., University of Salerno,\\
Via Ponte don Melillo, 84084 Fisciano (SA), Italy\\
e-mail: ciarlett@diima.unisa.it\\
e-mail: iovane@diima.unisa.it\\
and\\
\and M.\,A.\,Sumbatyan \\
Rostov State University, Faculty of Mechanics and Mathematics,\\
Zorge Street 5, Rostov-on-Don 344090, Russia\\
e-mail: sumbat@math.rsu.ru}
\date{}
\maketitle

\begin{abstract}
The paper deals with classical problem for cracks dislocated in a
certain very specific porous elastic material, described by a
Cowin-Nunziato model. We propose a method based upon a reducing of
stress concentration problem for cracks to some integral
equations. By applying Fourier integral transforms the problem is
reduced to some integral equations. For the plane-strain problem
we operate with a direct numerical treatment of a hypersingular
integral equation. In the axially symmetric case, for the
penny-shaped crack, the problem is reduced to a regular Fredholm
integral equation of the second kind. In the both cases we study
stress-concentration factor, and investigate its behavior versus
porosity of the material. More in particular the stress
concentration factor in the medium with voids is always higher,
under the same conditions, than in the classical elastic medium
made of material of the skeleton. Further, as can be seen, the
influence of the porosity becomes more significant for larger
cracks; that is also quite natural from a physical point of view.

\end{abstract}

\section{Introduction}

There are a number of theories about mechanical properties of
porous materials. One of them is a Biot consolidation theory of
fluid-saturated porous solids [1]. Typically, these theories
reduce to classical elasticity when the pore fluid is absent. This
is why Cowin and Nunziato proposed a new theory to describe
properties of homogeneous elastic materials with voids free of
fluid [2]. \ This theory is a special case of microstretch
elasticity of Eringen, when micropolar effects are discarded.
Eringen's microstretch theory is more appropriate for geological
materials like rocks, soils since, this theory takes into account
the intrinsic rotations and stretch of materials \cite{eringen},
\cite{eringen2}. These effects are not considered in
Cowin-Nunziato theory. However, the last theory is mathematically
simpler to deal with, and it takes into account voids in porous
materials without any other phase, like liquid and gas.

A general theory of such materials is currently well-developed by
many authors (see, for example [3],[4]), but too few concrete
problems are solved, to allow one to estimate the practical merits
of this model.

Generally, this theory is founded on the balance of energy, where presence
of the pores involves additional degree of freedom, namely, the fraction of
elementary volume. As a consequence, the bulk mass density is given by the
product of two fields, the void volume fraction and the mass density of the
matrix (elastic) material.

An exact explicit solution is well known in the problem for a
crack (both linear and penny-shaped) dislocated in the classical
linear elastic space [9]. If a normal load is applied to the faces
of the crack, then the shape of the faces near the edge of the
crack under this stress can be represented explicitly as a
root-square function. This permits analytical calculation of the
stress concentration coefficient in the classical case. Obviously,
stress concentration analysis is also very important in
engineering practice for porous materials. Therefore, the main
goal of the present work is to construct a strict solution of the
static crack problem for the line (plane-strain problem) and penny
shaped (axially symmetric problem) cracks dislocated in the linear
elastic granular (porous) space.

We first give a short summary of basic equations, then demonstrate
application of the Fourier transform that allows us to reduce the problem to
some integral equations and to construct a direct numerical collocation
technique to solve this equation. In the plane-strain problem we treat
numerically a certain hypersingular integral equation, and a special kind of
the collocation technique can be also applied to such equations. Finally, we
demonstrate in figures how the calculated stress concentration coefficient
at the crack edge depends upon variation of some physical and geometric
parameters.

\section{Governing Equations for Elastic Media with Voids: Plane-Strain
Problem}

We refer the deformation of the continuum to a fixed system of rectangular
Cartesian coordinates $Oxyz$. Let us consider elastic material with voids
which possesses a reference configuration with a constant volume fraction $%
\nu_0$. The considered theory asserts that the constant mass density $%
\varrho $ has the decomposition [2] $\varrho=\gamma\nu$, where $\gamma$ is
the density of the matrix material, and $\nu$ ~($0<\nu\le 1$) is the volume
fraction field.

Let $\phi $ ($\phi =\nu -\nu _{0}$) be the change in volume
fraction from the reference one. Then the linear theory of
homogeneous and isotropic elastic material with voids is described
by the following system of partial differential equations [2,8]
$$
\left\{
\begin{array}{c}
\mu ~\Delta ~\bar{u}+(\lambda +\mu )~{\rm grad~div}~\bar{u}+\beta ~{\rm grad}%
~\phi =0\vspace*{3mm} \\
\alpha ~\Delta ~\phi -\xi ~\phi -\beta ~{\rm div}~\bar{u}=0\quad ,
\end{array}
\right. \eqno(2.1)
$$
where $\mu $ and $\lambda $ are classical elastic constants; $\alpha ,~\beta
$ and $\xi $ -- some constants related to porosity of the medium. Besides, $%
\bar{u}$ denotes the displacement vector. Obviously, if $\beta =0$, then the
elastic and the ``porosity'' fields are independent. Thus, in the case $%
\beta =0$ the stress-strain state is insensitive to the function
$\phi $. The components of the stress tensor are defined, in terms
of the functions $\bar{u}$ and $\phi $, by the following relations
($\delta _{ij}$ is the Kronecker's delta)
$$
\left\{
\begin{array}{l}
\sigma _{ij}=\lambda ~\delta _{ij}~\varepsilon _{kk}+2~\mu ~\varepsilon
_{ij}+\beta ~\phi ~\delta _{ij} \\
\varepsilon _{ij}=\displaystyle\frac{1}{2}(u_{i,j}+u_{j,i})~.
\end{array}
\right. \eqno(2.2)
$$

Let us formulate the plane-strain boundary value problem for the
porous (granular) medium. In this case $\bar{u}=\{u_{x}(x,y),~u_{y}(x,y),~0%
\} $, and the basic system (2.1) can be rewritten as follows
$$
\left\{
\begin{array}{l}
\displaystyle\frac{\partial ^{2}u_{x}}{\partial x^{2}}+c^{2}\,\frac{\partial
^{2}u_{x}}{\partial y^{2}}+(1-c^{2})~\frac{\partial ^{2}u_{y}}{\partial
x\partial y}+H~\frac{\partial \phi }{\partial x}=0\vspace*{3mm} \\
\displaystyle\frac{\partial ^{2}u_{y}}{\partial y^{2}}+c^{2}\,\frac{\partial
^{2}u_{y}}{\partial x^{2}}+(1-c^{2})~\frac{\partial ^{2}u_{x}}{\partial
x\partial y}+H~\frac{\partial \phi }{\partial y}=0\vspace*{3mm} \\
\displaystyle l_{1}^{2}\left( \frac{\partial ^{2}\phi }{\partial x^{2}}+%
\frac{\partial ^{2}\phi }{\partial y^{2}}\right) -\frac{l_{1}^{2}}{l_{2}^{2}}%
~\phi -\left( \frac{\partial u_{x}}{\partial x}+\frac{\partial u_{y}}{%
\partial y}\right) =0\quad ,
\end{array}
\right. \eqno(2.3)
$$
with
$$
\begin{array}{c}
\displaystyle\frac{\sigma _{xx}}{\lambda +2\mu }=\frac{\partial u_{x}}{%
\partial x}+(1-2c^{2})~\frac{\partial u_{y}}{\partial y}+H\phi
~,\vspace*{3mm} \\
\displaystyle\frac{\sigma _{yy}}{\lambda +2\mu }=(1-2c^{2})~\frac{\partial
u_{x}}{\partial x}+\frac{\partial u_{y}}{\partial y}+H\phi ~,\vspace*{3mm}
\\
\displaystyle\frac{\sigma _{xy}}{\mu }=\frac{\partial u_{x}}{\partial y}+%
\frac{\partial u_{y}}{\partial x}\quad .
\end{array}
\eqno(2.4)
$$
The positive physical parameters (2.3)-(2.4) are introduced as follows
$$
c^{2}=\frac{\mu }{\lambda +2\mu }~,\quad H=\frac{\beta }{\lambda +2\mu }%
~,\quad l_{1}^{2}=\frac{\alpha }{\beta }~,\quad l_{2}^{2}=\frac{\alpha }{\xi
}~,\eqno(2.5)
$$
where the first two numbers $c~,~H$ are dimensionless and the quantities $%
l_{1}~,~l_{2}$ have the dimension of length.

Let us consider a thin crack of the length $2a$ with plane faces, dislocated
over the segment $-a<x<a$ along the $x$-axis. Let the plane-strain
deformation of this crack be caused by a constant stress $%
\sigma_{yy}=\sigma_0$ applied at infinity (i.e. at $y=\pm\infty$). Then, due
to linearity of the problem, it is easily seen that both the shape of the
crack's faces and the stress concentration at the crack's edges are the same
as in the problem with a solution decaying at infinity and the following
boundary conditions corresponding to the case when the normal load $%
-\sigma_0 $ is symmetrically applied to the faces of the crack and there is
no load at infinity.

For the last problem the boundary conditions over the line $y=0$ are
$$
\sigma_{xy}=0~,~\frac{\partial\phi}{\partial y}=0~(|x|<\infty)~,\quad
\sigma_{yy}=-\sigma_0~(|x|<a)~,\quad u_y=0~(|x|>a)~. \eqno(2.6)
$$
It is proved (cf.[7,10]) that the boundary condition for the
function $\phi$ (2.6) directly follows from a balance principle.

Let us apply the Fourier transform along the x-axis to relations (2.3),(2.4)
and the boundary conditions (2.6). Then Eqs.(2.3) reduce to the following
system of ordinary differential equations, with respect to images of the
functions $u_x,\,u_y,\,\phi$. Therefore, these become functions of the
variable $y$ only, with the Fourier parameter $s$ instead of the first
variable $x$ (all Fourier images are denoted by respective capital letters,
for all physical quantities):
$$
\left\{
\begin{array}{l}
\displaystyle c^2
U^{\prime\prime}_x-s^2\,U_x+(1-c^2)(-is)\,U^{\prime}_y-is\,H\,\Phi=0
\vspace*{3mm} \\
\displaystyle(1-c^2)(-is)\,U^{\prime}_x +
U^{\prime\prime}_y-c^2\,s^2\,U_y+H\,\Phi^{\prime}=0 \vspace*{3mm} \\
\displaystyle is\,U_x-U^{\prime}_y+l_1^2\,\Phi^{\prime\prime}-\left(\frac{%
l_1^2}{l_2^2}+l_1^2\,s^2\right) \,\Phi=0~.
\end{array}
\right. \eqno(2.7)
$$

Let us introduce the new unknown function $g(x)$ as
$$
u_y(x,0)= \left\{
\begin{array}{c}
\displaystyle g(x)~,\quad |x|<a \vspace*{3mm} \\
\displaystyle 0~,\qquad |x|>a
\end{array}
~, \right. \eqno(2.8)
$$
then the boundary conditions (2.6) in Fourier images become
$$
U^{\prime}_x-is~U_y=0,\qquad \Phi^{\prime}=0,\qquad U_y=G(s)~,\qquad\quad
y=0~, \eqno(2.9a)
$$
where
$$
G(s)=\int\limits_{-a}^a g(\xi)\exp(is\xi)d\xi~. \eqno(2.9b)
$$

General solution of the system (2.7) is constructed in accordance with a
classical theory of ordinary differential equations (cf.[10]). Due to
natural symmetry of the problem with respect to $x$-axis, we give here only
solution for $y\ge 0$, which has the following form (here all physical
quantities of dimension of length, including components of the displacement
vector, are written in a dimensionless form being related to $l_{2}$)
$$
\begin{array}{c}
\left(
\begin{array}{c}
U_{x}\vspace*{3mm} \\
U_{y}\vspace*{3mm} \\
\Phi
\end{array}
\right) =D_{1}\left(
\begin{array}{c}
\displaystyle\frac{isH}{1-N}\vspace*{3mm} \\
\displaystyle\frac{Hq(s)}{1-N}\vspace*{3mm} \\
\displaystyle1
\end{array}
\right) \displaystyle e^{-q(s)y}~+~D_{2}\left(
\begin{array}{c}
\displaystyle i~sign(s)\vspace*{3mm} \\
\displaystyle1\vspace*{3mm} \\
\displaystyle0
\end{array}
\right) \displaystyle e^{-|s|y}~+\vspace*{4mm} \\
D_{3}\left(
\begin{array}{c}
\displaystyle\frac{1-N+c^{2}}{(-is)(N-1+c^{2})}+iy~sign(s)\vspace*{3mm} \\
\displaystyle y\vspace*{3mm} \\
\displaystyle\frac{2Nc^{2}}{H(1-N-c^{2})}
\end{array}
\right) \displaystyle e^{-|s|y},~q(s)=\sqrt{s^{2}+1-N},
\end{array}
\eqno(2.10)
$$
where $N=(l_{2}^{2}/l_{1}^{2})\,H~,\quad 0\leq N<1$\quad is the so-called
''coupling number''. Obviously, the unknown constants $D_{1},\,D_{2},\,D_{3}$
should be defined from boundary conditions (2.9a). These yield the 3 x 3
algebraic system
$$
\left\{
\begin{array}{l}
\displaystyle\frac{sHq(s)}{1-N}~D_{1}+s\,D_{2}+\frac{(1-N)\,sign(s)}{%
N-1+c^{2}}~D_{3}~=~0\vspace*{3mm} \\
\displaystyle q(s)~D_{1}~+~\frac{2Nc^{2}\,|s|}{H(1-N-c^{2})}%
~D_{3}~=~0\vspace*{3mm} \\
\displaystyle\frac{Hq(s)}{1-N}~D_{1}+D_{2}~=~G(s)~,
\end{array}
\right. \eqno(2.11)
$$
whose principal determinant is
$$
\Delta =\frac{(1-N)q(s)sign(s)}{N-1+c^{2}}~.\eqno(2.12a)
$$
The particular ones (related to the three unknowns when one
applies the Cramer's rule) are:
$$
\Delta _{1}=-G(s)\,\frac{2Nc^{2}s|s|}{H(N-1+c^2)}~,\eqno(2.12b)
$$
$$
\Delta _{2}=G(s)\,\frac{q(s)sign(s)[(1-N)^{2}+2Nc^{2}s^{2}]}{(1-N)(N-1+c^{2})%
}~,\eqno(2.12c)
$$
$$
\Delta _{3}=-G(s)q(s)s~.\eqno(2.12d)
$$
Thus, the Fourier image $P(s)$ of the normal stress $\sigma _{yy}$
over the line $y=0$ as follows
$$
\begin{array}{c}
\displaystyle\frac{P(s)}{2\mu }=\frac{Hs^{2}}{1-N}\,D_{1}+|s|\,D_{2}+\frac{%
c^{2}}{N-1+c^{2}}\,D_{3}=\vspace*{4mm} \\
\displaystyle=\frac{G(s)|s|}{(1-N)^{2}\,q(s)}\left[
2Nc^{2}s^{2}(q-|s|)+(1-N)(1-N-c^{2})q\right] ~.
\end{array}
\eqno(2.13)
$$

By applying Fourier inversion to relation (2.13), with the use of a
convolution theorem, one can express the normal stress on the line of
symmetry $y=0$ in terms of its opening:
$$
\frac{(1-N)^2}{2\mu}\sigma_{yy}(x,0)= \int\limits_{-b}^b
g(\xi)K(x-\xi)d\xi~,\quad \left(b=\frac{a}{l_2}\right)~, \eqno(2.14)
$$
where
$$
\begin{array}{c}
\displaystyle K(x)=\frac{1}{2\pi}\int\limits_{-\infty}^\infty\frac{|s|}{q(s)%
} [2Nc^2s^2(q-|s|)+(1-N)(1-N-c^2)q]e^{-isx}ds \vspace*{4mm} \\
\displaystyle =\frac{1}{\pi}\int\limits_{0}^\infty L(s)\cos(sx)ds,\qquad
q=q(s)=\sqrt{s^2+1-N}~, \vspace*{4mm} \\
\displaystyle L(s)=\frac{s}{q(s)}\left[2Nc^2s^2(q-s)+(1-N)(1-N-c^2)q\right]~.
\end{array}
\eqno(2.15)
$$

Finally, the boundary condition $\sigma _{yy}(x,0)=-\sigma
_{0}~,\quad (|x|<a)$ from (2.6) yields the main integral equation
for the considered plane-strain problem with the convolution
kernel
$$
\int\limits_{-b}^{b}g(\xi )K(x-\xi )d\xi =-\,(1-N)^{2}\,\frac{\sigma _{0}}{%
2\mu }~,\qquad |x|<b~.\eqno(2.16)
$$

\section{Reduction to Integral Equation in the Axially Symmetric Case}

In this case the general equations (2.1) reduce, instead of Eqs.(2.3), to
the following system of partial differential equations for the components of
the displacement vector in cylindrical coordinates $\bar
u=\{u_r(x,y),~0,~u_z(x,y)\}$
$$
\left\{
\begin{array}{l}
\displaystyle \left(\frac{\partial^2 u_r}{\partial r^2}+ \frac{1}{r}\frac{%
\partial u_r}{\partial r}- \frac{u_r}{r^2}\right)+c^2\,\frac{\partial^2 u_r}{%
\partial z^2}+ (1-c^2)~\frac{\partial^2 u_z}{\partial r\partial z}+ H~\frac{%
\partial\phi}{\partial r}=0 \vspace*{3mm} \\
\displaystyle (1-c^2)\,\frac{\partial}{\partial z} \left(\frac{\partial u_r}{%
\partial r}+\frac{u_r}{r}\right)+ c^2\,\left(\frac{\partial^2 u_z}{\partial
r^2}+ \frac{1}{r}\frac{\partial u_z}{\partial r}\right)+ \frac{\partial^2 u_z%
}{\partial z^2}+ H~\frac{\partial\phi}{\partial z}=0 \vspace*{3mm} \\
\displaystyle l_1^2\left(\frac{\partial^2\phi}{\partial r^2}+ \frac{1}{r}%
\frac{\partial\phi}{\partial r}+ \frac{\partial^2\phi}{\partial z^2}\right)-%
\frac{l_1^2}{l_2^2}~\phi- \left(\frac{\partial u_r}{\partial r}+\frac{u_r}{r}%
+ \frac{\partial u_z}{\partial z}\right) =0 \quad ,
\end{array}
\right. \eqno(3.1)
$$
with
$$
\begin{array}{c}
\displaystyle\frac{\sigma_{rr}}{\lambda+2\mu}=\frac{\partial u_r}{\partial r}%
+ (1-2c^2)~\left(\frac{u_r}{r}+ \frac{\partial u_z}{\partial z}%
\right)+H\phi~, \vspace*{3mm} \\
\displaystyle\frac{\sigma_{zz}}{\lambda+2\mu}= (1-2c^2)~\left(\frac{\partial
u_r}{\partial r}+\frac{u_r}{r}\right)+ \frac{\partial u_z}{\partial z}%
+H\phi~, \vspace*{3mm} \\
\displaystyle\frac{\sigma_{rz}}{\mu}=\frac{\partial u_r}{\partial z}+ \frac{%
\partial u_z}{\partial r}\quad,
\end{array}
\eqno(3.2)
$$

If a round thin plane crack is placed at the center of the cylindrical
coordinates: $z=0~,~r<a$ and its faces is under a normal constant load $%
-\sigma_0$, then the boundary conditions over the plane of symmetry $z=0$
are
$$
\sigma_{rz}=0~,~\frac{\partial\phi}{\partial z}=0~(r<\infty)~,\quad
\sigma_{zz}=-\sigma_0~(r<a)~,\quad u_z=0~(r>a)~. \eqno(3.3)
$$

Let us apply the Hankel transform along the $r$-variable to relations
(3.1),(3.2):
$$
U_r(s,z)=\int\limits_0^{\infty}u_r(r,z)\,J_1(sr)\,rdr~,\qquad
u_r(r,z)=\int\limits_0^{\infty}U_r(s,z)\,J_1(sr)\,sds~, \eqno(3.4a)
$$
$$
U_z(s,z)=\int\limits_0^{\infty}u_z(r,z)\,J_0(sr)\,rdr~,\qquad
u_z(r,z)=\int\limits_0^{\infty}U_z(s,z)\,J_0(sr)\,sds~, \eqno(3.4b)
$$
$$
\phi(s,z)=\int\limits_0^{\infty}\Phi(r,z)\,J_0(sr)\,rdr~,\qquad
\Phi(r,z)=\int\limits_0^{\infty}\phi(s,z)\,J_0(sr)\,sds~. \eqno(3.4c)
$$

If one applies the Hankel transform to equations (3.1), then one comes to
the system of ordinary differential equations, with respect to images of the
functions $u_r,\,u_z,\,\phi$. Therefore, these become functions of the
variable $z$ only, with some parameter $s$:
$$
\left\{
\begin{array}{l}
\displaystyle c^2
U^{\prime\prime}_r-s^2\,U_r-(1-c^2)\,s\,U^{\prime}_z-H\,s\,\Phi=0
\vspace*{3mm} \\
\displaystyle(1-c^2)\,s\,U^{\prime}_r +
U^{\prime\prime}_z-c^2\,s^2\,U_z+H\,\Phi^{\prime}=0 \vspace*{3mm} \\
\displaystyle -s\,U_r-U^{\prime}_z+l_1^2\,\Phi^{\prime\prime}-\left(\frac{%
l_1^2}{l_2^2}+l_1^2\,s^2\right) \,\Phi=0~,
\end{array}
\right. \eqno(3.5)
$$
where all derivatives are applied with respect to the variable $z$.

Solution of Eqs.(3.5), together with Eqs.(3.4) and boundary conditions (3.3)
gives for the Hankel image of the normal stress $\sigma_{zz}$ the following
expression in terms of the image of crack's faces opening
$$
\frac{P(s)}{\mu}=\frac{G(s)\,s}{(1-N)^2\,q(s)} \left[%
2Nc^2s^2(q-s)+(1-N)(1-N-c^2)q\right]~, \eqno(3.6)
$$
where
$$
u_z(r,0)= \left\{
\begin{array}{c}
\displaystyle g(r)~,\quad r<b \vspace*{3mm} \\
\displaystyle 0~,\qquad r>b
\end{array}
\right. \eqno(3.7)
$$
and
$$
G(s)=\int\limits_0^b g(\rho)\,J_0(s\rho)\rho\,d\rho~,\qquad
P(s)=\int\limits_0^\infty \sigma_{zz}(\rho,0)\,J_0(s\rho)\rho\,d\rho~. \eqno%
(3.8)
$$

If one applies inverse Hankel transform to relation (3.8) then one arrives
at the main integral equation for the round crack problem:
$$
\int\limits_0^b g(\rho)\rho K(r,\rho)d\rho=-(1-N)^2\,\frac{\sigma_0}{\mu}~,
\quad 0<r<b \quad (b=\frac{a}{l_2}) \eqno(3.9a)
$$
with the kernel
$$
K(r,\rho)=\int\limits_{0}^\infty L(s)\,J_0(rs)J_0(\rho s)\,s^2\,ds. \eqno%
(3.9b)
$$

\section{Properties of Integral Equations and Numerical Implementation for
the Plane Problem}

Let us start from the evident estimate
$$
q(s)~\sim ~s\left[ 1+\frac{1-N}{2s^{2}}+O(\frac{1}{s^{4}})\right] \quad {\rm %
at}\quad s\rightarrow +\infty ~,\eqno(4.1)
$$
so the asymptotic behavior of the symbolic function of the kernel at
infinity is given as follows
$$
L(s)~\sim ~(1-N)^{2}\,(1-c^{2})\,s+O\left( \frac{1}{s}\right) ~,\qquad
s\rightarrow \infty ~.\eqno(4.2)
$$
It is obvious that the leading asymptotic term (4.2) leads to a
principal degree of the kernel's (2.15) singularity, and one can
see that the latter is defined by the integral
$$
\int\limits_{0}^{\infty }\,s\cos (sx)ds=-\frac{1}{x^{2}}~,\eqno(4.3)
$$
i.e. the kernel is hypersingular when $x\rightarrow 0$ [9]. It can be shown
that the remaining term in asymptotic estimate (4.2) possesses a logarithmic
(i.e. weak) singularity at $x\rightarrow 0$.

However, the kernel (2.15) admits explicit representation that permits
direct estimate of its singular properties. To obtain such a representation,
we calculate the following integrals
$$
\int\limits_0^\infty\,s^3 \cos(sx)ds= -\,\frac{d^2}{dx^2}\int\limits_0^%
\infty\,s \cos(sx)ds= \frac{6}{x^4}~. \eqno(4.4)
$$
Then we consider the following table integral (here $K_0$ is a McDonald's
function of the order $0$)
$$
\int\limits_0^\infty\,\frac{\cos(sx)}{q(s)}ds=K_0(\sqrt{1-N}|x|)~, \eqno%
(4.5)
$$
and apply, step by step, sequential derivatives to Eq.(4.5), as follows
(assuming $x>0$)
$$
\begin{array}{c}
\displaystyle \int\limits_0^\infty\,\frac{s^2\,\cos(sx)}{q(s)}ds=-\,\frac{d^2%
}{dx^2} \left[K_0(\sqrt{1-N}\,x)\right]= \vspace*{4mm} \\
\displaystyle =\sqrt{1-N}\frac{d}{dx}K_1(\sqrt{1-N}x)= (1-N)\,\frac{d}{d(%
\sqrt{1-N}x)}K_1(\sqrt{1-N}x)= \vspace*{4mm} \\
\displaystyle = -\left[(1-N)\,K_0(\sqrt{1-N}\,x)+\frac{\sqrt{1-N}}{x}\,K_1(%
\sqrt{1-N}\,x)\right]~,
\end{array}
\eqno(4.6)
$$
$$
\begin{array}{c}
\displaystyle \int\limits_0^\infty\,\frac{s^4\,\cos(sx)ds}{(1-N)\,q(s)}=%
\frac{d^2}{dx^2} \left[K_0(\sqrt{1-N}\,x)+\frac{K_1(\sqrt{1-N}\,x)}{\sqrt{1-N%
}\,x}\right]= \vspace*{4mm} \\
\displaystyle =2\left(\frac{3}{x^3}+\frac{1-N}{x}\right) \frac{K_1(\sqrt{1-N}%
\,x)}{\sqrt{1-N}}+ \left(1-N+\frac{3}{x^2}\right) K_0(\sqrt{1-N}\,x),
\end{array}
\eqno(4.7)
$$
where we have used the derivatives of the McDonald's functions of the zero's
and the first order:
$$
\frac{dK_0(z)}{dz}=-K_1(z)~,\qquad \frac{dK_1(z)}{dz}=-K_0(z)-\frac{K_1(z)}{z%
}~. \eqno(4.8)
$$

Therefore, the full kernel for the plane-strain crack problem is explicitly
represented as follows
$$
\begin{array}{c}
\displaystyle K(x)=\frac{1}{\pi}\left\{2Nc^2\left[\frac{6}{x^4}- 2\sqrt{1-N}%
\left(\frac{3}{|x|^3}+\frac{1-N}{|x|}\right)K_1(\sqrt{1-N}%
\,|x|)-\right.\right. \vspace*{4mm} \\
\displaystyle \left.\left.-(1-N)\left(1-N+\frac{3}{x^2}\right)K_0(\sqrt{1-N}%
\,|x|)\right] -\frac{(1-N)(1-N-c^2)}{x^2}\right\}.
\end{array}
\eqno(4.9)
$$

>From the last representation it is obvious that the function $K(x)$ (4.9)
is differentiable as many times as wanted outside a small vicinity of the
origin $x=0$. Let us estimate its behavior when $x\to 0$. For this aim one
can use the following asymptotic formulas
$$
\begin{array}{c}
\displaystyle K_0(z)~\sim~-\ln\left(\frac{z}{2}\right)-\gamma+O(z^2\,\ln
z)~, \quad z\to +0~, \vspace*{3mm} \\
\displaystyle K_1(z)~\sim~\frac{1}{z}+\frac{z}{2}\ln\left(\frac{z}{2}\right)
+\left(\gamma-\frac{1}{2}\right)\frac{z}{2}+O(z^3\,\ln z)~, \quad z\to +0~,
\end{array}
\eqno(4.10)
$$
where $\gamma=0.577216$ is the Euler's constant [10]. Thus behavior of
function (4.9) for the small argument is
$$
K(x)~\sim~-\frac{(1-N)^2\,(1-c^2)}{\pi\, x^2}+O(\ln|x|)~, \quad x\to 0~, %
\eqno(4.11)
$$
so the kernel is hypersingular indeed.

The numerical method that we apply to solve the hypersingular
integral equation has been proposed in our previous paper [9]. It
is based on extraction of a characteristic hypersingular part of
the kernel. Let us represent the full equation as follows
$$
\int\limits_{-b}^{b}\left[ \frac{1}{\left( x-t\right) ^{2}}+K_{\ast }\left(
x,t\right) \right] g\left( t\right) dt=f^{\prime }\left( x\right) ,\quad
x\in \left( -b,b\right) ~,\eqno(4.12)
$$
where the regular part $K_{\ast }$ may admit a weak (i.e. integrable)
singularity. If we represent it as
$$
K_{\ast }\left( x,t\right) =\frac{\partial K_{1}\left( x,t\right) }{\partial
x},\eqno(4.13)
$$
then a bounded solution can be constructed by applying inversion of the
characteristic part, that reduces eq.(4.12) to a second-kind Fredholm
integral equation
$$
g\left( x\right) +\int\limits_{-b}^{b}N_{1}\left( x,t\right) g\left(
t\right) dt=f_{1}\left( x\right) ,\quad x\in \left( -b,b\right) ~,\eqno%
(4.14)
$$
where
$$
N_{1}\left( x,t\right) =\frac{\sqrt{b^{2}-x^{2}}}{\pi ^{2}}%
\int\limits_{-b}^{b}\frac{K_{1}\left( \tau ,t\right) d\tau }{\sqrt{%
b^{2}-\tau ^{2}}~\left( x-\tau \right) }~,\eqno(4.15a)
$$
$$
f_{1}\left( x\right) =\frac{\sqrt{b^{2}-x^{2}}}{\pi ^{2}}\int\limits_{-b}^{b}%
\frac{f\left( \tau \right) d\tau }{\sqrt{b^{2}-\tau ^{2}}~\left( x-\tau
\right) }~.\eqno(4.15a)
$$

Then we prove that, if $f\left( x\right) \in
C^{1}(-b,b);~K_{1}\left(
x,t\right) \in C^{1}\left[ \left( -b,b\right) \times \left( -b,b\right) %
\right] $, then for any $x\in \left( -b,b\right) $ the difference between
solution $g\left( x\right) $ of the linear algebraic system
$$
\sum\limits_{j=1}^{n}\left[ \frac{1}{x_{i}-t_{j}}-\frac{1}{x_{i}-t_{j-1}}%
+hK_{\ast }\left( x_{i},t_{j}\right) \right] \,g\left( t_{j}\right)
=f^{\prime }\left( x_{i}\right) ,~i=1,...,n\eqno(4.16)
$$
and the solution of eq.(4.14) tends to zero near the crack edge when $%
h\rightarrow 0$ (i.e. $n\rightarrow \infty $), where the mesh nodes $%
x_{i}=-b+\left( i-1/2\right) h\,,~i=1,...,n$ and $t_{j}=-b+jh,\quad
j=0,1...,n$ are taken with the constant step $h=2b/n$.

It should be noted that this result automatically implies: if our
numarically constructed solution of the system (4.16) tends to the
exact one, so the constructed solution tends to zero when
approaching the crack edges, as it follows from (4.14)-(4.15).

\section{Penny-Shaped Crack: Reduction to a Fredholm Integral Equation of
the Second Kind}

Here we apply a special transformation rather typical for the problems with
axial symmetry.

Let us rewrite Eq.(3.9b) in the following form
$$
\int\limits_0^b g(\rho)\rho d\rho \int\limits_{0}^\infty
L(s)\,J_0(rs)J_0(\rho s)\,s\,ds= -(1-N)^2\frac{\sigma_0}{\mu}~,\quad 0<r<b~, %
\eqno(5.1)
$$
and introduce the new function $\gamma(\xi)$ as follows
$$
\int\limits_{0}^b g(\rho)J_0(\rho s)\rho d\rho= \int\limits_{0}^\infty
g(\rho)J_0(\rho s)\rho d\rho= \int\limits_{0}^b \gamma(\xi)\frac{\sin(\xi s)%
}{s}d\xi~. \eqno(5.2)
$$

The necessary and sufficient condition for Eq.(5.2) to be correct is that
this must provide $g(r)=0$~ for $~r>b$. Let us control correctness of this
statement by applying inverse Hankel transform to (5.2):
$$
\begin{array}{c}
\displaystyle g(r)=\int\limits_{0}^\infty J_0(rs)ds\int\limits_{0}^b
\gamma(\xi)\sin(\xi s)d\xi= \vspace*{3mm} \\
\displaystyle =\int\limits_{0}^b \left\{
\begin{array}{r}
\displaystyle \frac{1}{\sqrt{\xi^2-r^2}}~,~r<\xi \\
\displaystyle 0 \quad~,~ r>\xi
\end{array}
\right\} d\xi= \left\{
\begin{array}{r}
\displaystyle \int\limits_r^b\frac{\gamma(\xi)d\xi}{\sqrt{\xi^2-r^2}}~,~r<b
\\
\displaystyle 0 \quad~,~ r>b
\end{array}
\right.~,
\end{array}
\eqno(5.3)
$$
so (5.2) is correct.

Now Eq.(5.1) becomes
$$
\int\limits_{0}^b \gamma(\xi)\sin(\xi s)d\xi \int\limits_{0}^\infty
L(s)\,J_0(rs)\,ds=-(1-N)^2\frac{\sigma_0}{\mu}~,\quad 0<r<b~, \eqno(5.4)
$$
and one may apply the operator
$$
Af=\int\limits_0^t\frac{rf(r)dr}{\sqrt{t^2-r^2}}~,\qquad 0<t<b \eqno(5.5)
$$
to the both sides of Eq.(5.4), that leads to the following integral equation
with respect to the function $\gamma(\xi)$
$$
\int\limits_{0}^b\gamma(\xi)d\xi\int\limits_{0}^\infty \frac{L(s)}{s}
\sin(\xi s)\sin(ts)ds=-(1-N)^2\frac{\sigma_0\,t}{\mu}~,\quad 0<t<b~, \eqno%
(5.6)
$$
if the applied normal load $\sigma_0$ is constant. When performing all above
transformations, we have used the table integrals
$$
\int\limits_{0}^\infty J_0(rs)\sin(\xi s)ds= \left\{
\begin{array}{r}
\displaystyle \frac{1}{\sqrt{\xi^2-r^2}}~,~r<\xi \\
\displaystyle 0 \quad~,~ r>\xi
\end{array}
\right., \quad \int\limits_0^t\frac{rJ_0(rs)dr}{\sqrt{t^2-r^2}}= \frac{%
\sin(ts)}{s}. \eqno(5.7)
$$

Let us extend the uknown function $\gamma(\xi)$ in Eq.(5.6) to a negative
interval $\xi\in(-b,0)$ as an odd function. Then the integral (5.6) can be
represented as follows ($\delta(x)$ is the Dirac's delta)
$$
\begin{array}{c}
\displaystyle\int\limits_{0}^b\gamma(\xi)d\xi\int\limits_{0}^\infty \frac{%
L(s)}{s} \sin(\xi s)\sin(ts)ds= \vspace*{3mm} \\
\displaystyle = \frac{1}{2}\int\limits_{0}^b\gamma(\xi)d\xi\int\limits_{0}^%
\infty L(s) \frac{\cos[(\xi-t) s]-\cos[(\xi+t)s]}{s}\,ds= \vspace*{3mm} \\
\displaystyle = \frac{1}{2}\int\limits_{0}^b\gamma(\xi)d\xi\int\limits_{0}^%
\infty \frac{L(s)}{s} \cos[(\xi-t) s]\,ds= \vspace*{3mm} \\
\displaystyle = \frac{1}{2}\int\limits_{0}^b\gamma(\xi)d\xi\int\limits_{0}^%
\infty \left[\frac{L(s)}{s}-(1-N)^2(1-c^2)\right] \cos[(\xi-t)s]\,ds+
\vspace*{3mm} \\
\displaystyle +\frac{\pi}{2}(1-N)^2(1-c^2)\gamma(t),\qquad {\rm since} \quad
\int\limits_0^\infty\cos(s\xi)ds=\pi\delta(\xi)~.
\end{array}
\eqno(5.8)
$$

Therefore, in this axially symmetric problem one arrives at a very regular
Fredholm integral equation of the second kind with the convolution kernel:
$$
\pi\gamma(t)+\int\limits_{-b}^b\gamma(\xi)K^*(t-\xi)d\xi= -\frac{2\sigma_0\,t%
}{(1-c^2)\mu}~,\quad |t|<b~, \eqno(5.9a)
$$
where
$$
\begin{array}{c}
\displaystyle K^*(x)=\int\limits_0^\infty \left[\frac{L(s)}{(1-N)^2(1-c^2)\,s%
}-1\right]\,\cos(xs)\,ds= \vspace*{3mm} \\
\displaystyle = \int\limits_0^\infty \left[\frac{%
2Nc^2s^2(q-s)+(1-N)(1-N-c^2)q(s)} {(1-N)^2(1-c^2)\,q(s)}-1\right]\cos(sx)ds.
\end{array}
\eqno(5.9b)
$$

Since the expression in the square brackets here is of the order $O(1/s^2)$ at $%
s\to\infty$ (see (4.2)), it can be easily proved that the kernel
$K^*(x)$ is regular. More precisely, it is differentiable over any
finite interval, in particular $K^*(x)\in C^1(-2b, 2b)$.

\section{Calculation of the Stress Concentration Coefficient}

Here we are intersted in behavior of the normal stress (which is $%
\sigma_{yy} $ for in-plane problem and $\sigma_{zz}$ for axially symmetric
problem) near the crack edge.

Let us start from the problem with axial symmetry. We first notice that
function $\gamma(\xi)$ is regular on the interval $\xi\in (-b,b)$, as a
solution of the regular Fredholm equation of the second kind (5.9). Then we
operate with the relation (5.4), which obviously determines the normal
stress $\sigma(r,0)$ not only for $0<r<b$, but also for $r>b$. It is easily
seen that with $r\to b+0$ the leading asymptotic term is given by the
leading term of $L(s)$ at $s\to\infty$:
\[
\begin{array}{c}
\displaystyle \frac{\sigma_{zz}(r,0)}{(1-c^2)\mu} \sim
\int\limits_0^b\sin(\xi s)d\xi \int\limits_0^\infty s J_0(rs)ds =
-\int\limits_0^b \gamma(\xi)\frac{d}{d\xi}\int\limits_0^\infty \cos(\xi
s)J_0(rs)ds= \vspace*{3mm} \\
\displaystyle = -\int\limits_0^b \gamma(\xi)\frac{d}{d\xi} \left\{
\begin{array}{l}
\displaystyle \frac{1}{\sqrt{r^2-\xi^2}}~,~\xi<r \\
\displaystyle \qquad\quad 0~,~\xi>r
\end{array}
\right\} d\xi=\int\limits_0^b \gamma(\xi)\frac{\xi d\xi}{\sqrt{r^2-\xi^2}^3}%
=\qquad\qquad (6.1) \vspace*{3mm} \\
\displaystyle =\int\limits_0^b \frac{\gamma(\xi)-\gamma(b)}{\sqrt{r^2-\xi^2}%
^3}\xi d\xi+ \frac{\gamma(b)}{\sqrt{r^2-b^2}}\quad\sim\quad \frac{\gamma(b)}{%
\sqrt{r^2-b^2}}~ \quad{\rm at}~\quad r\to b+0~,
\end{array}
\]
so the stress concentration coefficient is
$$
k=\lim\limits_{r\to b+0}\,\frac{|\sigma_{zz}(r,0)|}{\mu b(1-c^2)} \sqrt{%
r^2-b^2}\,=\,\frac{|\gamma(b)|}{b}~. \eqno(6.2)
$$

Some examples of calculation of this dimensionless coefficient for
the penny-shaped crack are shown in Figures 1 and 2.

In the plane-strain problem our approach is absolutely different, since
solution of the integral equation (2.16) with the hypersingular kernel (4.9)
is constructed numerically from the algebraic system (4.16), so that the
applied numerical method does not operate explicitly with the root-square
structure $\sqrt{b^2-x^2}$, vanishing at the edges of the crack. Therefore,
we need to treat expression (2.14), which represents the normal stress $%
\sigma_{yy}(x,0)$ also for $x>b$, in a direct numerical way.

Since $K(x)=-(1-N)^2\,(1-c^2)/(\pi x^2)+O(\ln|x|)~,~x\to 0$, it is obvious
that with $x\to b+0$ the leading asymptotic term of the considered normal
stress is
$$
\frac{\sigma_{yy}(x,0)}{2\mu(1-c^2)}=\frac{1}{(1-N)^2(1-c^2)}
\int\limits_{-b}^b g(\xi)K(x-\xi)d\xi~ \sim ~ -\frac{1}{\pi}%
\int\limits_{-b}^b \frac{g(\xi)d\xi}{(x-\xi)^2}, \eqno(6.3)
$$
so here the stress concentration factor is
$$
k=\lim\limits_{x\to b+0}\,\frac{|\sigma_{yy}(x,0)|}{\mu b(1-c^2)} \sqrt{%
x^2-b^2}\,= \frac{2}{\pi}\lim\limits_{x\to b+0} \left|\int\limits_{-b}^b
\frac{g(\xi)d\xi}{(x-\xi)^2}\right|\sqrt{x^2-b^2}\,. \eqno(6.4)
$$

Figures 3 and 4 demonstrate numerical results on computation of this
dimensionless constant for some values of parameters $N,\,c^2,\,b=a/l_2$.

\section{Conclusions}

\quad 1. We have proposed a method based upon a reducing of stress
concentration problem for cracks to some integral equations. In the
plane-strain problem this is a hypersingular integral equation, which
permits efficient direct numerical treatment. In the axially symmetric
problem for penny-shaped crack, after some traditional transformations, we
arrive at more regular Fredholm integral equation of the second kind. The
both types of equations can generally be solved numerically.

2. In some cases the developed equations admit exact analytical solution in
explicit form. The first case is for $N=0$, that is a classical linear
elastic material. In this case the plane-strain problem's integral equation
has only the characteristic component $-(1-N)^2\,(1-c^2)/(\pi x^2)$ and the
regular part $K_*$ vanishes. As can be seen from Eqs.(4.15) here $N_1\equiv
0 $, so exact solution is given by (4.14),(4.15b), that is obviously
coincides with a known classical solution. For the penny-shaped crack, the
kernel $K^*$ also vanishes, as can be directly seen from $Eq.(5.9b)$, so
function $\gamma(t)$ can be explicitly extracted from Eq.(5.9a), that
finally also leads to a well-known classical solution for the round plane
crack.

Another simple limiting case is when parameter $b=a/l_2$ is small, that
means the crack size to be small when compared with the physical parameter $%
l_2$ (the latter being of dimension of length). Indeed, in the axially
symmetric problem, if $b \to 0$, then integral operator in Eq.(5.9a)
vanishes. The same property takes place for the plane linear crack, since
integral of the regular part of the kernel in (4.12) vanishes. Thus, in the
both cases the stress-strain state near the small crack in the porous space
is like in an ideally elastic classical medium, that is quite natural from
the physical point of view.

3. If we investigate the influence of the porosity to the stress
concentration factor shown in figures, we can discover very interesting
properties. First of all, this factor in the medium with voids is always
higher, under the same conditions, than in the classical elastic medium made
of material of the skeleton. This can be explained with energetic arguments,
since in porous media a stress, distributed near the crack edge only in the
skeleton, can provide a balance of energy caused by applied normal force. So
this requires more intensity of the internal stress to provide the balance.

Further, as can be seen, influence of the porosity becomes more
significant for larger cracks; this is also quite natural from a
physical point of view. This results is very interesting
conclusion. Let us imagine a relatively small crack, which by a
sufficiently high stress concentration factor extends slightly in
its length. Then, as clear from the figures, the larger crack is
coupled with higher stress concentration factor that causes
further crack extension. This is the real physical mechanics of
cracks growth in porous media.

\newpage

\begin{center}
{\Large {\bf LEGENDS TO FIGURES }}
\end{center}

\vspace*{1.5cm} Fig 1.~Relative value of the stress concentration factor $k$
with respect to $k_0$ in classical elastic medium versus coupling number,
penny-shaped crack: $c^2=0.2$~.

\vspace*{1.5cm} Fig 2.~Relative value of the stress concentration factor $k$
with respect to $k_0$ in classical elastic medium versus coupling number,
penny-shaped crack: $c^2=0.4$~.

\vspace*{1.5cm} Fig 3.~Relative value of the stress concentration factor $k$
with respect to $k_0$ in classical elastic medium versus coupling number,
plane linear crack: $c^2=0.2$~.

\vspace*{1.5cm} Fig 4.~Relative value of the stress concentration factor $k$
with respect to $k_0$ in classical elastic medium versus coupling number,
plane linear crack: $c^2=0.4$~.

\newpage \thispagestyle{empty} \qquad \qquad {\Large
\begin{picture}(300,200)

\thinlines
\put(0,0){\line(1,0){320}}
\put(0,200){\line(1,0){320}}
\put(0,0){\line(0,1){200}}
\put(320,0){\line(0,1){200}}
\put(80,0){\line(0,1){8}}
\put(160,0){\line(0,1){8}}
\put(240,0){\line(0,1){8}}

\put(0,40){\line(1,0){6}}
\put(-25,-2){$1.0$}
\put(-25,36){$1.2$}
\put(0,80){\line(1,0){6}}
\put(-25,76){$1.4$}
\put(0,120){\line(1,0){6}}
\put(-25,116){$1.6$}
\put(0,160){\line(1,0){6}}
\put(-25,156){$1.8$}
\put(-25,196){$2.0$}
\put(4,180){$k/k_0$}

{\large
\put(260,10){$b=1$}
\put(282,39){$b=5$}
\put(278,158){$b=10$}
}

\put(0,-18){$0$}
\put(70,-18){$0.2$}
\put(150,-18){$0.4$}
\put(230,-18){$0.6$}
\put(310,-18){$0.8$}
\put(270,-18){$N$}

\thicklines
\put(  0,   0){\line(  1,  0){ 10}}
\put( 10,   0){\line(  6,  1){ 10}}
\put( 20,   1){\line(  1,  0){ 10}}
\put( 30,   1){\line(  6,  1){ 10}}
\put( 40,   3){\line(  1,  0){ 10}}
\put( 50,   3){\line(  6,  1){ 10}}
\put( 60,   5){\line(  6,  1){ 10}}
\put( 70,   6){\line(  1,  0){ 10}}
\put( 80,   6){\line(  6,  1){ 10}}
\put( 90,   8){\line(  6,  1){ 10}}
\put(100,  10){\line(  6,  1){ 10}}
\put(110,  11){\line(  6,  1){ 10}}
\put(120,  13){\line(  6,  1){ 10}}
\put(130,  15){\line(  6,  1){ 10}}
\put(140,  17){\line(  6,  1){ 10}}
\put(150,  18){\line(  6,  1){ 10}}
\put(160,  20){\line(  6,  1){ 10}}
\put(170,  21){\line(  6,  1){ 10}}
\put(180,  23){\line(  4,  1){ 10}}
\put(190,  25){\line(  3,  1){ 10}}
\put(200,  29){\line(  4,  1){ 10}}
\put(210,  31){\line(  2,  1){ 20}}
\put(230,  41){\line(  5,  3){ 10}}
\put(240,  47){\line(  5,  3){ 10}}
\put(250,  53){\line(  4,  3){ 10}}
\put(260,  60){\line(  1,  1){ 10}}
\put(270,  70){\line(  4,  5){ 10}}
\put(280,  83){\line(  2,  3){ 10}}
\put(290,  98){\line(  2,  5){ 10}}
\put(300, 123){\line(  1,  3){ 10}}

\put(  0,   0){\line(  1,  0){ 10}}
\put( 10,   0){\line(  6,  1){ 10}}
\put( 20,   1){\line(  1,  0){ 10}}
\put( 30,   1){\line(  1,  0){ 10}}
\put( 40,   1){\line(  6,  1){ 10}}
\put( 50,   3){\line(  1,  0){ 10}}
\put( 60,   3){\line(  6,  1){ 10}}
\put( 70,   5){\line(  1,  0){ 10}}
\put( 80,   5){\line(  6,  1){ 10}}
\put( 90,   6){\line(  1,  0){ 10}}
\put(100,   6){\line(  6,  1){ 10}}
\put(110,   8){\line(  1,  0){ 10}}
\put(120,   8){\line(  6,  1){ 10}}
\put(130,  10){\line(  6,  1){ 10}}
\put(140,  11){\line(  6,  1){ 10}}
\put(150,  13){\line(  1,  0){ 10}}
\put(160,  13){\line(  5,  1){ 10}}
\put(170,  15){\line(  6,  1){ 10}}
\put(180,  17){\line(  3,  1){ 10}}
\put(190,  20){\line(  5,  1){ 10}}
\put(200,  22){\line(  5,  2){ 10}}
\put(210,  26){\line(  6,  1){ 10}}
\put(220,  28){\line(  5,  1){ 10}}
\put(230,  30){\line(  3,  1){ 10}}
\put(240,  33){\line(  6,  1){ 10}}
\put(250,  35){\line(  5,  2){ 10}}
\put(260,  39){\line(  2,  1){ 10}}
\put(270,  44){\line(  2,  1){ 10}}
\put(280,  49){\line(  3,  2){ 10}}
\put(290,  55){\line(  6,  5){ 10}}
\put(300,  64){\line(  1,  1){ 10}}

\put(  0,   0){\line(  1,  0){ 10}}
\put( 10,   0){\line(  1,  0){ 10}}
\put( 20,   0){\line(  1,  0){ 10}}
\put( 30,   0){\line(  1,  0){ 10}}
\put( 40,   0){\line(  1,  0){ 10}}
\put( 50,   0){\line(  6,  1){ 10}}
\put( 60,   1){\line(  1,  0){ 10}}
\put( 70,   1){\line(  1,  0){ 10}}
\put( 80,   1){\line(  1,  0){ 10}}
\put( 90,   1){\line(  1,  0){ 10}}
\put(100,   1){\line(  1,  0){ 10}}
\put(110,   1){\line(  1,  0){ 10}}
\put(120,   1){\line(  1,  0){ 10}}
\put(130,   1){\line(  1,  0){ 10}}
\put(140,   1){\line(  6,  1){ 10}}
\put(150,   3){\line(  1,  0){ 10}}
\put(160,   3){\line(  1,  0){ 10}}
\put(170,   3){\line(  1,  0){ 10}}
\put(180,   3){\line(  1,  0){ 10}}
\put(190,   3){\line(  1,  0){ 10}}
\put(200,   3){\line(  1,  0){ 10}}
\put(210,   3){\line(  1,  0){ 10}}
\put(220,   3){\line(  6,  1){ 10}}
\put(230,   5){\line(  1,  0){ 10}}
\put(240,   5){\line(  1,  0){ 10}}
\put(250,   5){\line(  1,  0){ 10}}
\put(260,   5){\line(  1,  0){ 10}}
\put(270,   5){\line(  6,  1){ 10}}
\put(280,   6){\line(  1,  0){ 10}}
\put(290,   6){\line(  1,  0){ 10}}
\put(300,   6){\line(  1,  0){ 10}}


\end{picture}
} {\Large \vspace*{8cm} }

\begin{center}
{\Large Fig.1 }
\end{center}

\newpage \vspace*{2cm} \thispagestyle{empty} \qquad \qquad {\Large
\begin{picture}(300,200)

\thinlines
\put(0,0){\line(1,0){300}}
\put(0,240){\line(1,0){300}}
\put(0,0){\line(0,1){240}}
\put(300,0){\line(0,1){240}}
\put(50,0){\line(0,1){8}}
\put(100,0){\line(0,1){8}}
\put(150,0){\line(0,1){8}}
\put(200,0){\line(0,1){8}}
\put(250,0){\line(0,1){8}}

\put(0,40){\line(1,0){6}}
\put(-25,-2){$1.0$}
\put(-25,36){$1.2$}
\put(0,80){\line(1,0){6}}
\put(-25,76){$1.4$}
\put(0,120){\line(1,0){6}}
\put(-25,116){$1.6$}
\put(0,160){\line(1,0){6}}
\put(-25,156){$1.8$}
\put(0,200){\line(1,0){6}}
\put(-25,196){$2.0$}
\put(-25,236){$2.2$}
\put(4,220){$k/k_0$}

{\large
\put(250,15){$b=1$}
\put(262,80){$b=5$}
\put(235,200){$b=10$}
}

\put(0,-18){$0$}
\put(90,-18){$0.2$}
\put(190,-18){$0.4$}
\put(290,-18){$0.6$}
\put(255,-18){$N$}

\thicklines
\put(  0,   0){\line(  6,  1){ 10}}
\put( 10,   1){\line(  6,  1){ 10}}
\put( 20,   3){\line(  6,  1){ 10}}
\put( 30,   5){\line(  6,  1){ 10}}
\put( 40,   6){\line(  6,  1){ 10}}
\put( 50,   8){\line(  5,  1){ 10}}
\put( 60,  10){\line(  6,  1){ 10}}
\put( 70,  12){\line(  6,  1){ 10}}
\put( 80,  13){\line(  5,  2){ 10}}
\put( 90,  17){\line(  5,  1){ 10}}
\put(100,  19){\line(  3,  1){ 10}}
\put(110,  23){\line(  4,  1){ 10}}
\put(120,  25){\line(  3,  2){ 10}}
\put(130,  32){\line(  5,  2){ 10}}
\put(140,  36){\line(  2,  1){ 10}}
\put(150,  41){\line(  2,  1){ 10}}
\put(160,  46){\line(  2,  1){ 20}}
\put(180,  57){\line(  2,  1){ 30}}
\put(210,  72){\line(  6,  5){ 10}}
\put(220,  81){\line(  1,  1){ 10}}
\put(230,  91){\line(  3,  4){ 10}}
\put(240, 105){\line(  3,  5){ 10}}
\put(250, 121){\line(  2,  3){ 10}}
\put(260, 136){\line(  1,  3){ 10}}
\put(270, 166){\line(  1,  6){ 10}}
\put(280, 226){\line(  1,  3){ 10}}

\put(  0,   0){\line(  6,  1){ 10}}
\put( 10,   1){\line(  1,  0){ 10}}
\put( 20,   1){\line(  5,  1){ 10}}
\put( 30,   3){\line(  6,  1){ 10}}
\put( 40,   5){\line(  6,  1){ 10}}
\put( 50,   7){\line(  6,  1){ 10}}
\put( 60,   8){\line(  6,  1){ 10}}
\put( 70,  10){\line(  6,  1){ 10}}
\put( 80,  11){\line(  6,  1){ 10}}
\put( 90,  13){\line(  6,  1){ 10}}
\put(100,  15){\line(  6,  1){ 10}}
\put(110,  17){\line(  4,  1){ 10}}
\put(120,  19){\line(  5,  1){ 10}}
\put(130,  21){\line(  4,  1){ 10}}
\put(140,  24){\line(  4,  1){ 10}}
\put(150,  26){\line(  3,  1){ 10}}
\put(160,  29){\line(  3,  1){ 10}}
\put(170,  33){\line(  5,  2){ 20}}
\put(190,  40){\line(  3,  1){ 10}}
\put(200,  43){\line(  2,  1){ 10}}
\put(210,  49){\line(  5,  3){ 10}}
\put(220,  55){\line(  1,  1){ 10}}
\put(230,  65){\line(  3,  2){ 10}}
\put(240,  71){\line(  1,  1){ 30}}
\put(270, 100){\line(  4,  5){ 10}}
\put(280, 112){\line(  2,  3){ 10}}

\put(  0,   0){\line(  1,  0){ 10}}
\put( 10,   0){\line(  1,  0){ 10}}
\put( 20,   0){\line(  6,  1){ 10}}
\put( 30,   1){\line(  1,  0){ 10}}
\put( 40,   1){\line(  1,  0){ 10}}
\put( 50,   1){\line(  1,  0){ 10}}
\put( 60,   1){\line(  1,  0){ 10}}
\put( 70,   1){\line(  6,  1){ 10}}
\put( 80,   3){\line(  1,  0){ 10}}
\put( 90,   3){\line(  1,  0){ 10}}
\put(100,   3){\line(  1,  0){ 10}}
\put(110,   3){\line(  6,  1){ 10}}
\put(120,   5){\line(  1,  0){ 10}}
\put(130,   5){\line(  1,  0){ 10}}
\put(140,   5){\line(  1,  0){ 10}}
\put(150,   5){\line(  1,  0){ 10}}
\put(160,   5){\line(  6,  1){ 10}}
\put(170,   6){\line(  1,  0){ 10}}
\put(180,   6){\line(  1,  0){ 10}}
\put(190,   6){\line(  6,  1){ 10}}
\put(200,   8){\line(  1,  0){ 10}}
\put(210,   8){\line(  1,  0){ 10}}
\put(220,   8){\line(  1,  0){ 10}}
\put(230,   8){\line(  6,  1){ 10}}
\put(240,  10){\line(  1,  0){ 10}}
\put(250,  10){\line(  1,  0){ 10}}
\put(260,  10){\line(  6,  1){ 10}}
\put(270,  11){\line(  1,  0){ 10}}
\put(280,  11){\line(  6,  1){ 10}}


\end{picture}
} {\Large \vspace*{8cm} }

\begin{center}
{\Large Fig.2 }
\end{center}

\newpage \thispagestyle{empty} \qquad \qquad {\Large
\begin{picture}(300,200)

\thinlines
\put(0,0){\line(1,0){320}}
\put(0,200){\line(1,0){320}}
\put(0,0){\line(0,1){200}}
\put(320,0){\line(0,1){200}}
\put(80,0){\line(0,1){8}}
\put(160,0){\line(0,1){8}}
\put(240,0){\line(0,1){8}}

\put(0,40){\line(1,0){6}}
\put(-25,-2){$1.0$}
\put(-25,36){$1.2$}
\put(0,80){\line(1,0){6}}
\put(-25,76){$1.4$}
\put(0,120){\line(1,0){6}}
\put(-25,116){$1.6$}
\put(0,160){\line(1,0){6}}
\put(-25,156){$1.8$}
\put(-25,196){$2.0$}
\put(4,180){$k/k_0$}

{\large
\put(250,17){$b=1$}
\put(282,34){$b=3$}
\put(278,145){$b=10$}
}

\put(0,-18){$0$}
\put(70,-18){$0.2$}
\put(150,-18){$0.4$}
\put(230,-18){$0.6$}
\put(310,-18){$0.8$}
\put(270,-18){$N$}

\thicklines
\put(  0,   0){\line(  1,  0){ 10}}
\put( 10,   0){\line(  1,  0){ 10}}
\put( 20,   0){\line(  6,  1){ 10}}
\put( 30,   1){\line(  1,  0){ 10}}
\put( 40,   1){\line(  1,  0){ 10}}
\put( 50,   1){\line(  1,  0){ 10}}
\put( 60,   1){\line(  6,  1){ 10}}
\put( 70,   3){\line(  1,  0){ 10}}
\put( 80,   3){\line(  1,  0){ 10}}
\put( 90,   3){\line(  6,  1){ 10}}
\put(100,   5){\line(  1,  0){ 10}}
\put(110,   5){\line(  6,  1){ 10}}
\put(120,   6){\line(  1,  0){ 10}}
\put(130,   6){\line(  6,  1){ 10}}
\put(140,   8){\line(  6,  1){ 10}}
\put(150,  10){\line(  1,  0){ 10}}
\put(160,  10){\line(  4,  1){ 10}}
\put(170,  12){\line(  6,  1){ 10}}
\put(180,  14){\line(  5,  1){ 10}}
\put(190,  16){\line(  5,  1){ 10}}
\put(200,  18){\line(  4,  1){ 10}}
\put(210,  20){\line(  3,  1){ 10}}
\put(220,  23){\line(  5,  2){ 10}}
\put(230,  28){\line(  5,  2){ 10}}
\put(240,  31){\line(  5,  3){ 10}}
\put(250,  37){\line(  3,  2){ 10}}
\put(260,  44){\line(  6,  5){ 10}}
\put(270,  52){\line(  1,  1){ 10}}
\put(280,  62){\line(  3,  5){ 10}}
\put(290,  79){\line(  1,  2){ 10}}
\put(300,  99){\line(  1,  4){ 10}}

\put(  0,   0){\line(  1,  0){ 10}}
\put( 10,   0){\line(  1,  0){ 10}}
\put( 20,   0){\line(  1,  0){ 10}}
\put( 30,   0){\line(  6,  1){ 10}}
\put( 40,   1){\line(  1,  0){ 10}}
\put( 50,   1){\line(  1,  0){ 10}}
\put( 60,   1){\line(  1,  0){ 10}}
\put( 70,   1){\line(  1,  0){ 10}}
\put( 80,   1){\line(  6,  1){ 10}}
\put( 90,   3){\line(  1,  0){ 10}}
\put(100,   3){\line(  1,  0){ 10}}
\put(110,   3){\line(  1,  0){ 10}}
\put(120,   3){\line(  6,  1){ 10}}
\put(130,   5){\line(  1,  0){ 10}}
\put(140,   5){\line(  1,  0){ 10}}
\put(150,   5){\line(  1,  0){ 10}}
\put(160,   5){\line(  6,  1){ 10}}
\put(170,   6){\line(  1,  0){ 10}}
\put(180,   6){\line(  1,  0){ 10}}
\put(190,   6){\line(  6,  1){ 10}}
\put(200,   8){\line(  1,  0){ 10}}
\put(210,   8){\line(  1,  0){ 10}}
\put(220,   8){\line(  6,  1){ 10}}
\put(230,  10){\line(  1,  0){ 10}}
\put(240,  10){\line(  6,  1){ 10}}
\put(250,  11){\line(  1,  0){ 10}}
\put(260,  11){\line(  6,  1){ 10}}
\put(270,  13){\line(  1,  0){ 10}}
\put(280,  13){\line(  6,  1){ 10}}
\put(290,  15){\line(  6,  1){ 10}}
\put(300,  16){\line(  6,  1){ 10}}

\put(  0,   0){\line(  1,  0){ 10}}
\put( 10,   0){\line(  6,  1){ 10}}
\put( 20,   1){\line(  1,  0){ 10}}
\put( 30,   1){\line(  1,  0){ 10}}
\put( 40,   1){\line(  6,  1){ 10}}
\put( 50,   3){\line(  1,  0){ 10}}
\put( 60,   3){\line(  1,  0){ 10}}
\put( 70,   3){\line(  6,  1){ 10}}
\put( 80,   5){\line(  1,  0){ 10}}
\put( 90,   5){\line(  6,  1){ 10}}
\put(100,   6){\line(  1,  0){ 10}}
\put(110,   6){\line(  6,  1){ 10}}
\put(120,   8){\line(  6,  1){ 10}}
\put(130,  10){\line(  1,  0){ 10}}
\put(140,  10){\line(  6,  1){ 10}}
\put(150,  11){\line(  6,  1){ 10}}
\put(160,  13){\line(  1,  0){ 10}}
\put(170,  13){\line(  4,  1){ 10}}
\put(180,  15){\line(  6,  1){ 10}}
\put(190,  17){\line(  6,  1){ 10}}
\put(200,  19){\line(  5,  1){ 10}}
\put(210,  21){\line(  4,  1){ 10}}
\put(220,  23){\line(  4,  1){ 10}}
\put(230,  26){\line(  4,  1){ 10}}
\put(240,  28){\line(  3,  1){ 10}}
\put(250,  32){\line(  5,  2){ 10}}
\put(260,  36){\line(  5,  2){ 10}}
\put(270,  40){\line(  2,  1){ 10}}
\put(280,  45){\line(  3,  2){ 10}}
\put(290,  51){\line(  4,  3){ 10}}
\put(300,  59){\line(  6,  5){ 10}}

\end{picture}
} {\Large \vspace*{8cm} }

\begin{center}
{\Large Fig.3 }
\end{center}

\newpage \vspace*{2cm} \thispagestyle{empty} \qquad \qquad {\Large
\begin{picture}(300,200)

\thinlines
\put(0,0){\line(1,0){300}}
\put(0,200){\line(1,0){300}}
\put(0,0){\line(0,1){200}}
\put(300,0){\line(0,1){200}}
\put(50,0){\line(0,1){8}}
\put(100,0){\line(0,1){8}}
\put(150,0){\line(0,1){8}}
\put(200,0){\line(0,1){8}}

\put(0,40){\line(1,0){6}}
\put(-25,-2){$1.0$}
\put(-25,36){$1.2$}
\put(0,80){\line(1,0){6}}
\put(-25,76){$1.4$}
\put(0,120){\line(1,0){6}}
\put(-25,116){$1.6$}
\put(0,160){\line(1,0){6}}
\put(-25,156){$1.8$}
\put(0,200){\line(1,0){6}}
\put(-25,196){$2.0$}
\put(4,180){$k/k_0$}

{\large
\put(250,28){$b=1$}
\put(262,62){$b=3$}
\put(250,150){$b=6$}
}

\put(0,-18){$0$}
\put(90,-18){$0.2$}
\put(190,-18){$0.4$}
\put(290,-18){$0.6$}
\put(255,-18){$N$}

\thicklines
\put(  0,   0){\line(  6,  1){ 10}}
\put( 10,   1){\line(  1,  0){ 10}}
\put( 20,   1){\line(  6,  1){ 10}}
\put( 30,   3){\line(  6,  1){ 10}}
\put( 40,   5){\line(  1,  0){ 10}}
\put( 50,   5){\line(  5,  1){ 10}}
\put( 60,   7){\line(  6,  1){ 10}}
\put( 70,   8){\line(  6,  1){ 10}}
\put( 80,  10){\line(  6,  1){ 10}}
\put( 90,  11){\line(  6,  1){ 10}}
\put(100,  13){\line(  5,  1){ 10}}
\put(110,  15){\line(  5,  1){ 10}}
\put(120,  17){\line(  4,  1){ 10}}
\put(130,  20){\line(  4,  1){ 10}}
\put(140,  22){\line(  3,  1){ 10}}
\put(150,  26){\line(  4,  1){ 10}}
\put(160,  28){\line(  5,  2){ 10}}
\put(170,  32){\line(  5,  2){ 10}}
\put(180,  36){\line(  5,  2){ 10}}
\put(190,  40){\line(  2,  1){ 10}}
\put(200,  45){\line(  5,  3){ 10}}
\put(210,  51){\line(  5,  3){ 10}}
\put(220,  57){\line(  5,  4){ 10}}
\put(230,  65){\line(  6,  5){ 10}}
\put(240,  73){\line(  1,  1){ 10}}
\put(250,  83){\line(  3,  4){ 10}}
\put(260,  97){\line(  3,  5){ 10}}
\put(270, 113){\line(  1,  2){ 10}}
\put(280, 133){\line(  2,  5){ 10}}

\put(  0,   0){\line(  6,  1){ 10}}
\put( 10,   1){\line(  1,  0){ 10}}
\put( 20,   1){\line(  6,  1){ 10}}
\put( 30,   3){\line(  6,  1){ 10}}
\put( 40,   5){\line(  1,  0){ 10}}
\put( 50,   5){\line(  5,  1){ 10}}
\put( 60,   7){\line(  6,  1){ 10}}
\put( 70,   8){\line(  6,  1){ 10}}
\put( 80,  10){\line(  6,  1){ 10}}
\put( 90,  11){\line(  6,  1){ 10}}
\put(100,  13){\line(  6,  1){ 10}}
\put(110,  15){\line(  5,  1){ 10}}
\put(120,  17){\line(  5,  1){ 10}}
\put(130,  19){\line(  4,  1){ 10}}
\put(140,  21){\line(  4,  1){ 10}}
\put(150,  24){\line(  4,  1){ 10}}
\put(160,  26){\line(  4,  1){ 10}}
\put(170,  29){\line(  3,  1){ 10}}
\put(180,  32){\line(  3,  1){ 10}}
\put(190,  36){\line(  5,  2){ 10}}
\put(200,  40){\line(  5,  2){ 10}}
\put(210,  43){\line(  2,  1){ 10}}
\put(220,  49){\line(  2,  1){ 10}}
\put(230,  54){\line(  5,  3){ 10}}
\put(240,  60){\line(  5,  3){ 10}}
\put(250,  66){\line(  4,  3){ 10}}
\put(260,  73){\line(  6,  5){ 10}}
\put(270,  81){\line(  1,  1){ 10}}
\put(280,  91){\line(  5,  6){ 10}}

\put(  0,   0){\line(  1,  0){ 10}}
\put( 10,   0){\line(  6,  1){ 10}}
\put( 20,   1){\line(  1,  0){ 10}}
\put( 30,   1){\line(  1,  0){ 10}}
\put( 40,   1){\line(  6,  1){ 10}}
\put( 50,   3){\line(  1,  0){ 10}}
\put( 60,   3){\line(  6,  1){ 10}}
\put( 70,   5){\line(  1,  0){ 10}}
\put( 80,   5){\line(  1,  0){ 10}}
\put( 90,   5){\line(  6,  1){ 10}}
\put(100,   6){\line(  1,  0){ 10}}
\put(110,   6){\line(  6,  1){ 10}}
\put(120,   8){\line(  1,  0){ 10}}
\put(130,   8){\line(  6,  1){ 10}}
\put(140,  10){\line(  1,  0){ 10}}
\put(150,  10){\line(  6,  1){ 10}}
\put(160,  11){\line(  6,  1){ 10}}
\put(170,  13){\line(  1,  0){ 10}}
\put(180,  13){\line(  6,  1){ 10}}
\put(190,  15){\line(  1,  0){ 10}}
\put(200,  15){\line(  5,  1){ 10}}
\put(210,  17){\line(  6,  1){ 10}}
\put(220,  18){\line(  1,  0){ 10}}
\put(230,  18){\line(  5,  1){ 10}}
\put(240,  20){\line(  6,  1){ 10}}
\put(250,  22){\line(  6,  1){ 10}}
\put(260,  23){\line(  6,  1){ 10}}
\put(270,  25){\line(  6,  1){ 10}}
\put(280,  27){\line(  6,  1){ 10}}

\end{picture}
} {\Large \vspace*{8cm} }

\begin{center}
{\Large Fig.4 }
\end{center}

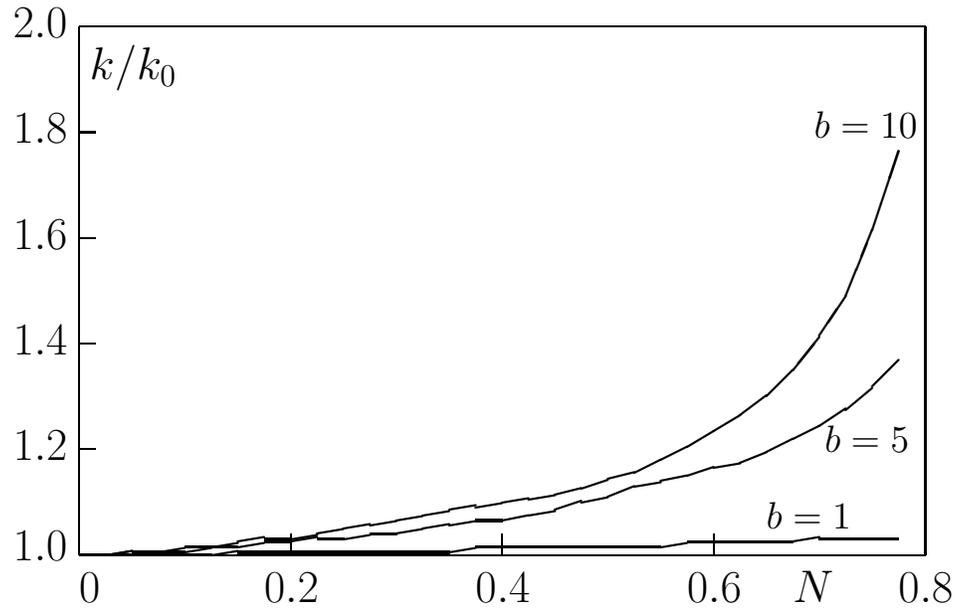
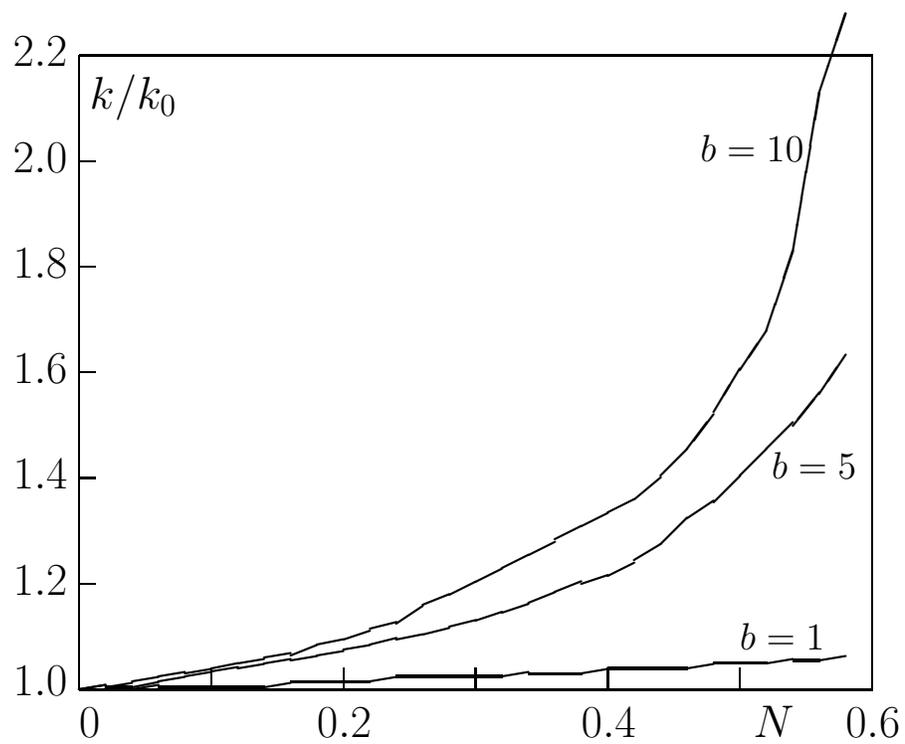
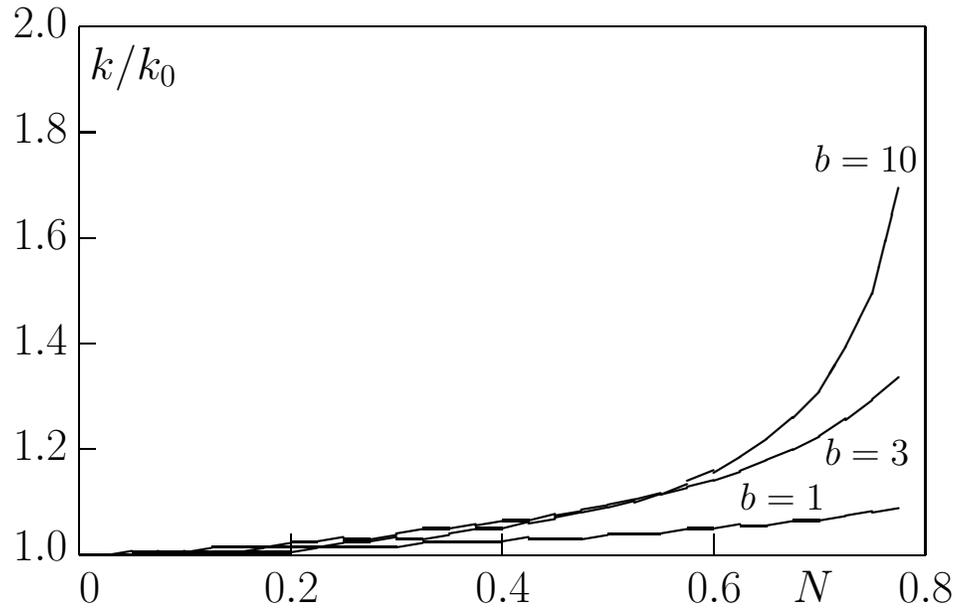
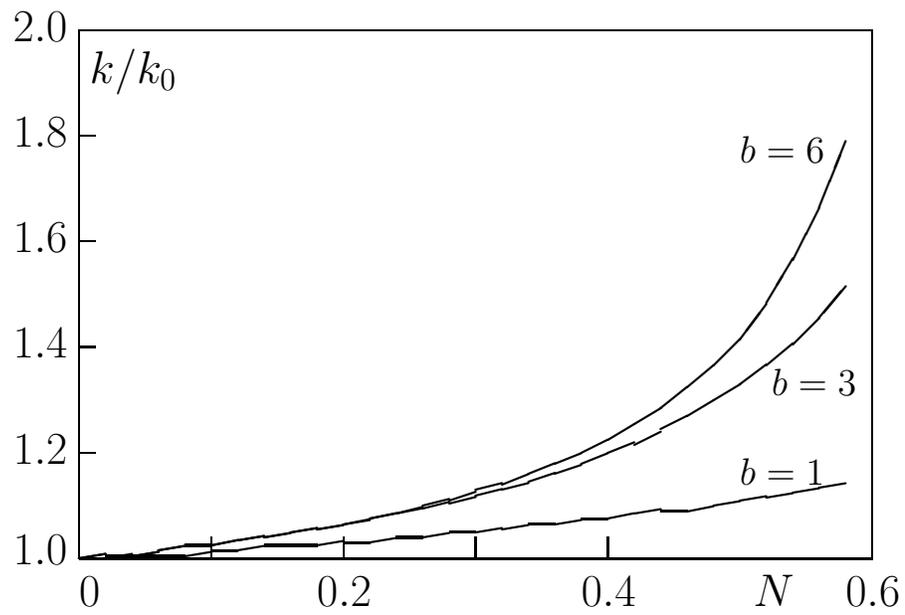
\end{document}